\documentclass[pra,aps,preprint]{revtex4-2}
\usepackage{epsf,epsfig}
\usepackage{amsmath}
\usepackage[colorlinks=true,linkcolor=blue,urlcolor=blue,citecolor=blue,pdfusetitle]{hyperref}
\usepackage[utf8]{inputenc}
\usepackage[english]{babel}
\usepackage{amsmath}
\usepackage[caption = false]{subfig}
\usepackage{graphicx,epstopdf}
\usepackage{blindtext}
\usepackage[table,xcdraw,dvipsnames]{xcolor}
\usepackage{lipsum}
\usepackage{amsfonts}
\usepackage{bbm}
\usepackage{amssymb}
\usepackage{enumerate}
\usepackage{color}
\usepackage{latexsym}
\usepackage{times,txfonts}
\usepackage{wrapfig}

\usepackage{orcidlink}

\begin{document}
\title{Quantum Speed Limit Time in two-qubit system by Dynamical Decoupling Method}

\author{Arefeh Aaliray }
%\email{Arefe.Aaliray@yahoo.com}
\affiliation{Faculty of Physics, University of Isfahan, Hezar Jarib, P. O. Box 81746-73441, Isfahan, Iran}

\author{Hamidreza Mohammadi\!\!~\orcidlink{0000-0001-7046-3818}}
\email{hr.mohammadi@sci.ui.ac.ir }
\affiliation{Faculty of Physics, University of Isfahan, Hezar Jarib, P. O. Box 81746-73441, Isfahan, Iran}
\affiliation{Quantum Optics Group, Faculty of Physics, University of Isfahan, Hezar Jarib, P. O. Box 81746-73441, Isfahan, Iran}
\begin{abstract}
\noindent Quantum state change can not occurs instantly, but the speed of quantum evolution is limited to an upper bound value, called quantum speed limit (QSL). Engineering QSL is an important task for quantum information and computation science and technologies. This paper devotes to engineering QSL and quantum correlation in simple two-qubit system suffering dephasing via Periodic Dynamical Decoupling (PDD) method in both Markovian and non-Markovian dynamical regimes. The results show that when decoupling pulses are applied to both qubits this method removes all undesirable effects of the dephasing process, completely. Applying the PDD on only one of the qubits also works but with lower efficiency. Additionally, ultra-high speedup of the quantum processes become possible during the pulse application period, for enough large number of  pulses. The results is useful for high speed quantum gate implementation application.
\end{abstract}
\maketitle
\section{Introduction}
The laws of nature impose an unavoidable limit on the speed of physical processes. In quantum systems this limitation stems from uncertainty principle and is called by Quantum Speed Limit (QSL). Maximum speed of dynamical evolution in quantum system is determined by Quantum Speed Limit Time(QSLT); minimum time required for a quantum system to evolve between two distinguishable quantum states\cite{zhang.srep2014}.  The notion of  QSLT plays important roles in different scenarios, including quantum communication\cite{zhang.srep1-bekenstein1981energy}, the identification of precision bounds in quantum metrology\cite{zhang.srep2-giovannetti2011advances}, the formulation of computational limits of physical systems\cite{zhang.srep3-lloyd2002computational}, as well as the development of quantum optimal control algorithms\cite{zhang.srep4-caneva2009optimal}. Mathematically, for a closed system with unitary evolution there are unified lower bounds of QSLT called Mandelstam-Tamm (MT) type bound and Margolus-Levitin (ML) type bound
\cite{zhang.srep5-mandelstam1945energy,zhang.srep6-fleming1973unitarity,
zhang.srep7-anandan1990geometry,zhang.srep8-vaidman1992minimum, zhang.srep9-margolus1998maximum,zhang.srep10-levitin2009fundamental}. The extensions of the MT and ML bounds to non-orthogonal states and to driven systems have been investigated in Refs.\cite{zhang.srep11-PhysRevA.67.052109,zhang.srep12-PhysRevA.82.022107,zhang.srep13-Deffner_2013,zhang.srep14-PhysRevLett.70.3365,zhang.srep15-RevModPhys.67.759}. The QSLT for non-unitary evolution of open systems is also studied\cite{zhang.srep16-taddei2013quantum,zhang.srep17-del2013quantum,zhang.srep18-deffner2013quantum}.These bounds are related to the concept of the passage time $\tau_p$, which is the required time for a given pure state $|\chi\rangle$ to become orthogonal to itself under unitary dynamics\cite{campo4-schulman2008jump} Mandelstam and Tamm showed that the passage time can be lower bounded by the inverse of the variance in the energy of the system $\Delta E=(\langle E^2\rangle -\langle E\rangle^2)^\frac12$, so that
\begin{equation}\label{1}
\tau_{MT}\geq\frac{\pi\hbar}{2\Delta E},
\end{equation}
whenever the dynamics under study is governed by a Hermitian Hamiltonian\cite{campo13-zwierz2012comment}. While Margolus and Levitin found that the lower bound of the passage time is reciprocal to the time-averaged of energy, $E$, and derived quantum speed limit time as:
\begin{equation}\label{2}
\tau_{ML}\geq \frac{\pi\hbar}{2E},
\end{equation}
 Combining Eqs.(1) and (2), obtained
\begin{equation}\label{3}
\tau_{QSL}=\max\{\tau_{MT},\tau_{ML}\}=\max\{\frac{\pi\hbar}{2\Delta E},\frac{\pi\hbar}{2E}\}.
\end{equation}
In ref.\cite{zhang.srep2014}, Zhang et al. have obtained the following expression for quantum speed limit bound for arbitrary initial states in the open system based on the von Neumann trace inequality and the Cauchy-Schwarz inequality:
\begin{equation}\label{QSLT}
\tau_{QSL}=\max\{\frac{1}{\langle\sum_{i=1}^n\sigma_i\rho_i\rangle_t},\frac{1}{\langle\sqrt{\sum_{i=1}^n\sigma_i^2}\rangle_t}\}\ast|f(\tau+\tau_d)-1| tr(\rho_\tau^2),
\end{equation}
where $\tau_d$ is driving time and $\sigma_i$ and $\rho_i$ are the singular values of $\dot{\rho}_t=L_t\rho_t$ and the initial state $\rho_\tau$ respectively, $\langle X \rangle_t=\tau_d^{-1}\int_{\tau}^{\tau+\tau_d}X\,dt$. The relative purity is considered as a distance measure to derive lower bound on the QSLT for open quantum systems. The relative purity between initial and final states of the quantum system is defined as $f(\tau+\tau_d)=tr[\rho_{\tau+\tau_d}\rho_\tau]/tr(\rho_\tau^{2})$. The first fraction in eq. (\ref{QSLT}) expresses a ML-type and the second one discusses a MT-type bound on the speed of quantum evolution valid for mixed initial states.

Furthermore, the main obstacle for implementing practical quantum information and computation tasks is decoherence; undesired leakage of the quantum coherence from the quantum systems, to their surrounding environment. To overcome this obstacle, it is necessary that the gating time, $\tau_G$ (the time needed for writing, processing, reading and sending the quantum information,) must be shorter than decoherence time, $\tau_{dec}$ (the time spent up to destroy superposition and hence quantum correlations lying on the state of the quantum system.) This condition ($\tau_G\ll\tau_{dec}$) requires ultra-fast quantum devices, which are very expensive and far from access. Indeed the decoherence time is related to the concept of Quantum Speed Limit Time (QSLT) which determines the maximum speed of dynamical evolution theoretically\cite{zhang.srep1-bekenstein1981energy}. A better solution is to postpone the decoherence by stretching the decoherence time. In this way, we need to control the speed of information exchanging between the quantum system and its environment. Here, a question has been raised: how one can delay and control this information exchange speed? The system-environment information exchanging and its rate are dependent to the system-environment interaction Hamiltonian. A pretty and effective solution is suppressing the system-environment interaction by continuously monitoring the system state via applying a sequence of pulses on it. This method is called the Dynamical Decoupling (DD) method. The DD techniques are the most successful methods to suppress decoherence in open qubit systems\cite{addis1-viola1998,addis2-viola1999, YIN2019136, howard2023implementing, barthel2022robust,PhysRevA.105.042614, PhysRevLett.128.230502}. The idea of the DD comes from the efforts had been done for eliminating the dephasing effects in the Nuclear Magnetic Resonance(NMR) spectroscopy by applying clever magnetic pulses on each nuclei\cite{slichter2013}. Mathematically, the idea of the DD phenomenon is reminiscent of the quantum Zeno effect\cite{viola20-presilla1996}. Among the various DD schemes the so-called bang-bang Periodic Dynamical Decoupling (PDD)\cite{addis2-viola1999} and its time-symmetrized version\cite{addis36-clausen2010bath} are interesting for us. As the performance of all DD schemes crucially determined by competition between on the time scale of the environment correlation function (memory time of environment) and characteristic time of external control field (pulse width and duty cycle). Indeed,  the decoherence process, may be modified if a time-varying control field acts on the dynamics of the system over time scales that are comparable to the memory time of the environment. The first efforts are began since the discovery of spin echoes in 1950\cite{viola22-hahn1950spin}, trying to eliminate the dephasing in NMR experiments. Then, Viola et. al. studied an exact model for a two-state quantum system (qubit) coupled to a thermal bath of harmonic oscillators. Their results showed that the quantum coherence of the system can be preserved by controlling the reservoir correlation time\cite{viola1998dynamical}. In this way, Uhrig has sought for a general solution to keep a qubit alive by employing optimized $\pi$-pulse sequences\cite{uhrig2007keeping-addis60}. The efficiency of DD method to suppress dephasing effects is examined in the ref.\cite{addis11-hodgson2010towards}, for different types of environments properties (Ohmic, sub-Ohmic and super-Ohmic). Influence of non-Markovianity of the dynamics on the effectiveness of a dynamical decoupling (DD) protocol for qubits undergoing pure dephasing is investigated in\cite{addis2015dynamical-47song}. The fidelity improvement using dynamical decoupling with superconducting qubits is studied by Pokharel et al.\cite{pokharel2018demonstration}. Engineering the QSLT via of DD protocol is investigated by Song et al\cite{song2017control}. They studied the variation of QSLT of a single qubit, embedded in a zero-temperature Ohmic-like dephasing reservoir, under PDD pulses. Their results reveal that QSLT is determined by initial and final quantum coherence of the qubit, as well as the non-Markovianity of the dynamics. It was shown that PDD pulses can modulate the final quantum coherence of the qubit and the amount of non-Markovianity. The results showed that for arbitrary initial states of the dephasing qubit with non-vanishing quantum coherence, PDD pulses can be used to induce potential acceleration of the quantum evolution in the short-time regime, while PDD pulses can lead to potential speedup and slowdown in the long-time regime. They demonstrated that the effect of PDD on the QSLT for the Ohmic or sub-Ohmic spectrum (Markovian reservoir) is much different from that for the super-Ohmic spectrum (non-Markovian reservoir).

All of these studies focus on the single-qubit systems and effects of D.D on the dynamics of quantum correlations and QSLT in the composite system have not considered yet. In this paper we consider a simple two-qubit system suffered by dephasing process and investigate dynamical behavior of quantum entanglement measured by concurrence and quantum discord and also QSLT when the interaction of one or both qubits with environment switched ON/OFF by DD method, respectively. The results show that the PDD pulses tend to recover the initial state of the system and hence all of the feature of the initial state such as the quantum entanglement and quantum discord and so on are preserved/recovered if the enough pulse number is applied to both qubits. Also, applying the pulses only on one of the qubits improves the situation but not completely. Furthermore, ultra-high speedup is possible during the time interval which PDD applied for both Markovian and non-Markovian dynamical regimes.

The paper is organized as follows. In section 2 the system of interest, namely a two-qubit system evolved under the pure dephasing including its exact solution in presence of PDD is introduced. In section 3,  quantum concurrence, quantum discord and quantum speed limit of this system is investigated before and after applying PDD. Finally, section 4 summarizes the findings and concludes the results.

\section{The model and solution}

The system under consideration includes a non-interacting two-qubit system embedded in surrounding environment. The inter-qubit separation is supposed large enough such that each qubit is coupled to a separate environment. The environments are modeled by bosonic bathes which are coupled to the qubits through pure-dephasing coupling, separately. Pure quantum dephasing process only araised by vacuum fluctuation of the reservoirs; hence to avoid the thermal effects, it is assumed that the reservoirs are held at zero-temperature i.e. the dephasing reservoir is initially in the vacuum state $\rho_B=|0\rangle_B\langle0|$\cite{song2017control}. Hence, the dynamics of each qubit is governed by the following single-qubit Hamiltonian ($\hbar=1$):
 \begin{equation}\label{10}
H_1=\frac{\omega_0}{2}\sigma_z+\sum_k\omega_k a_k^\dag a_k+\sum_k\sigma_z(g_k a_k^\dag+g_k^\ast a_k),
\end{equation}
where $\omega_0$ is the qubit frequency, $\sigma_{i=x,y,z}$ are the Pauli operators, $a_k^\dag(a_k)$ is the creation (annihilation) operator of the $k$th reservoir mode with the frequency $\omega_k$, and $g_k$ is the coupling constant associated with the qubit-$k$th reservoir mode interaction. Solving the master equation $\dot{\rho}_1(t)=\frac{1}{2} \dot{\Gamma}(t)[\sigma_z \rho_1(t)\sigma_z-\rho_1(t)]$, or employing the Kraus super-operator sum representation defined by $\rho_1(t)=\sum_{i=0}^1  E_i \rho_1(0) E_i^\dagger$ with following Kraus operators\cite{Haseli_2020},
\begin{eqnarray}
E_1=\sqrt{\frac{1+\gamma_0(t)}{2}} I_2, \,\,\,\,  E_2= \sqrt{\frac{1-\gamma_0(t)}{2}} \sigma_z,
\end{eqnarray}
yields the elements of the evolved single-qubit density operator as:
\begin{eqnarray}
\rho_1(t)=\left(\begin{array}{*{20}c}
\rho_{11}(0)&\rho_{12} \gamma_0(t)\\
\rho_{12}^* \gamma_0(t)& \rho_{22}(0)
\end{array}\right).
\end{eqnarray}

Where $I_2$ is $2 \times 2$ identity matrix and $\gamma_0(t)=e^{-\Gamma_0(t)}$ with the free (uncontrolled) decoherence function defined by:
\begin{equation}\label{Gamma0}
\Gamma_0(t)=\int_0^\infty\frac{I(\omega)}{\omega^2}[1-\cos(\omega t)]d\omega.
\end{equation}
Here $I(\omega)=\sum_j \delta(\omega-\omega_j)|g_j|^2$ is the spectral density function which encapsulates the behavior of the reservoir-qubit interaction. In this paper we consider the well studied class of spectral densities of the form\cite{addis28-leggett1987dynamics}:
\begin{equation}\label{31}
I(\omega)=\eta \frac{\omega^s}{\omega_c^{s-1}} e^{-\omega/\omega_c},
\end{equation}
where $s$ is the parameter of Ohmicity, $\eta$ a dimensionless coupling constant and $\omega_c$ a cut off frequency. Ohmic spectrum corresponds to $s=1$, while super-Ohmic spectra corresponds to $s>1$ and sub-Ohmic to $s<1$, respectively. The expression for $\Gamma_0(t)$ can be calculated analytically for both ohmic and non-ohmic cases\cite{song23-chin2012quantum}. For super and sub-ohmic cases the calculation gives:
\begin{equation}\label{32}
\Gamma_0(t, s\neq1)=\eta\Gamma(s-1)\{1-\frac{\cos[(s-1)\arctan(\omega_c, t)]}{(1+\omega_c^2 t^2)^{(s-1)/2}}\},
\end{equation}
where $\Gamma(s-1)$ being the Euler's Gamma function defined as $\Gamma(s-1)=\int_0^\infty d\omega  e^{-\omega} \omega^{s-2}$. Taking the limit $s\rightarrow1$, one also finds the $\Gamma_0(t)$ for Ohmic spectrum\cite{addis59-chin2012quantum,addis77-addis2014coherence}
as:
\begin{equation}\label{34}
\Gamma_0(t, s=1)=\frac{1}{2}\eta\ln(1+\omega_c^2 t^2).
\end{equation}
Under a train of ideal PDD $\pi$-pulses, the total single-Hamiltonian is switched to $H_1(t)=H_1+H_{DD}(t)$ with\cite{song2017control}:
\begin{equation}\label{10p}
H_{DD}(t)=\frac{\pi}{2} \sum_{j=0}^n\delta(t-\tau_j)\sigma_x.
\end{equation}
After switching $H_{DD}(t)$ ON, the dynamics of the qubit is still be exactly described by replacing $\Gamma_0(t)$ in above equations with  a modified (controlled) decoherence function $\Gamma(t)$\cite{addis11-hodgson2010towards} defined as following. Suppose that the total number of pulses, $N$, are applied at instants $\{t_1, ... t_n, ... t_f\}$, with $0<t_1<t_2<...<t_f<\tau_d$\cite{addis2015dynamical-47song} during an arbitrary storage time, $\tau_d$.
As shown by Uhrig\cite{addis61-uhrig2008exact}, the controlled decoherence function $\Gamma(t)$ can be written as
\begin{equation}\label{Gammat}
\Gamma(t)=
\begin{cases}
\Gamma_0(t)& t\leq t_1\\
\Gamma_{\#n}(t)& t_n<t\leq t_{n+1}, 0<n<N\\
\Gamma_{\#N}(t) & t\geq t_f,
\end{cases}
\end{equation}
for total pulse number, $N$ and
\begin{eqnarray}\label{35}
\Gamma_{\#n}(t) &=& (-1)^n\Gamma_0(t)+2\sum_{j=1}^{n}(-1)^{j+n}\Gamma_0 (t-\tau_j) \\ \nonumber &+& 2\sum_{j=1}^{n}(-1)^{j+1}\Gamma_0 (\tau_j)
+4\sum_{j=2}^{n}\sum_{k=1}^{j-1}(-1)^{j+k+1}\Gamma_0 (\tau_j-\tau_k),
\end{eqnarray}
is decoherence function after $n$th pulse. It is obvious that $\Gamma(t)=\Gamma_0(t)$ in the absence of PDD.

Generalizing the solution for the case of  two independent qubits each interacting with its own reservoir, for any arbitrary initial state, can be achieved by employing the recipe presented by Bellomo\cite{Bellomo-PhysRevLett.99.160502}. Following this recipe, the elements of two-qubit density matrix in the absence and/or presence of PDD on each qubit are derived as follow:
%\begin{widetext}
\begin{eqnarray}
\rho(t)=\left(\begin{array}{*{20}c}
\rho_{11}(0)&\rho_{12} P^1(t)&\rho_{13}(0) P^2(t)&\rho_{14}(0) P^1(t) P^2(t)\\
\rho_{12}^* P^1(t)& \rho_{22}(0)&\rho_{23}(0) P^1(t) P^2(t)&\rho_{24}(0) P^1(t)\\
\rho_{13}^*(0) P^2(t)&\rho_{23}^*(0) P^1(t) P^2(t)&\rho_{33}(0)&\rho_{34}(0) P^2(t)\\
\rho_{14}^*(0) P^1(t) P^2(t)&\rho_{24}^*(0) P^1(t)&\rho_{34}^*(0) P^2(t)&\rho_{44}(0)
\end{array}\right).
\end{eqnarray}
%\end{widetext}
With $\rho_{ij}(0)$s are the elements of the initial state and $
P^{\,\mu}(t)= e^{\Gamma(t)}$  is exponent of the harnessed decoherence function acting on the $\mu$th qubit ($\mu \in \{1,2\}$).

The equally spaced PDD $\pi$-pulses are applied at instants $\tau_n=(n\tau_f)(N+1)$ with $n=1,2,3,...,N$, and $\tau_f$ being the PDD pulses stop time. In fact, in practice, it is impossible to put infinite PDD pulses on the qubit i.e. $N\longrightarrow \infty$, and  also the errors due to pulse imperfection in real world are accumulated more as the number of pulses increases. Consequently,  we assume that the PDD pulses are applied during the finite time $\tau_f$, after which the system is subjected to the usual decoherence arising from its unavoidable environment. Knowing the dynamics of density matrix enables the calculation of the quantum features of the system, such as the entanglement, quantum discord, quantum consonance and QSLT for both short-term regime, i.e. $\tau_d=\tau_f$ and long-term regime, i.e. $\tau_d>\tau_f$. By fixing the actual driving time $\tau_d$, the term $\tau_{QSL}=\tau_d$ means that the quantum evolution possesses no potential capacity for further acceleration. While for the case of $\tau_{QSL}<\tau_d$, the shorter $\tau_{QSL}$ indicates the greater speedup potential capacity\cite{song13-xu2014quantum,song14-liu2015quantum}.

For an important class of initial states called X-states\cite{PhysRevA.77.042309,EPJD2011,QIP2017}, the density matrix at time $t$ for non-vanishing $\rho_{14}(0)$ and $\rho_{23}(0)$ can be expressed as:
\begin{eqnarray}
\rho(t)=\left(\begin{array}{*{20}c}
\rho_{11}(0)&0)&0&\rho_{14}(0) Q(t)\\
0& \rho_{22}(0)&\rho_{23}(0) Q(t)&0\\
0&\rho_{23}^*(0) Q(t)&\rho_{33}(0)&0\\
\rho_{14}^*(0) Q(t)&0&0&\rho_{44}(0)
\end{array}\right),
\end{eqnarray}
where $Q(t)=P^1(t) P^2(t)$ .
The concurrence of this state, $C(\rho(t))=C_t$ can be written as:
\begin{equation}
C_t=C_0 |Q(t)|,
\end{equation}
here  $C_0=C(\rho(0))= 2 \max\{0, |\rho_{14}(0)|-\sqrt{\rho_{22}(0) \rho_{33}(0)}, |\rho_{23}(0)|-\sqrt{\rho_{11}(0) \rho_{44}(0)}\}$ is the concurrence of initial state.
Quantum consonance of this state is given by:
\begin{equation}
QC_t=QC_0 Q(t),
\end{equation}
where  $QC_0=QC(\rho(0))= 2(Re(\rho_{14}(0)+\rho_{23}(0)))$ is the quantum consonance of the initial state.
For the special case of the singlet state as initial state, it is easy to show that quantum discord of the state at time $t$ is:
\begin{equation}
QD_t=QD(\rho(t))=1-H(\frac{1+|Q(t)|}{2}),
\end{equation}
where $H(x)=-x \log(x)-(1-x)\log(1-x)$ is the Shannon entropy.
For a general X-shaped initial density matrix suffered by pure dephasing process, the QSLT can be calculated by following formula:
\begin{equation}
\frac{\tau_{QSL}(t)}{\tau_d}=\Phi_0 \frac{|1-Q(t)|}{\int_{0}^{\tau_d} |\dot{Q}(t)| dt}.
\end{equation}
Where
\begin{equation}
\Phi_0=\max\{\frac{2(|\rho_{14}(0)|^2+|\rho_{23}(0)|^2)}{(\rho_{11}(0)+\rho_{44}(0))|\rho_{14}(0)|+(\rho_{22}(0)+\rho_{33}(0))|\rho_{23}(0)|},\sqrt{2(|\rho_{14}(0)|^2+|\rho_{23}(0)|^2)}\},
\end{equation}
depends only, on the parameters of initial state. Since $\int_{a}^{b} |f(x)|dx \geq |\int_{a}^{b} f(x)dx|$, one can obtain an upper bound for $\tau_{QSL}$ as:
\begin{equation}
\frac{\tau_{QSL}(t)}{\tau_d} \leq \Phi_0 \frac{1-e^{-\Gamma(t)}}{1-e^{-\Gamma(\tau_d)}}.
\end{equation}

Dependence of the quantum correlations ($C_t, QC_t$ and $QD_t$) on the function $Q(t)$ stems to the fact that these function are related to the coherence of the system and hence are affacted by decoherence. So, it is obvious that the function $Q(t)$ plays an essential role to determine quantum feature of the system, therefore, in the next section we first study the properties of $Q(t)$ and then its effects on the dynamics of QSL.

\begin{figure}
\centering
\includegraphics[width=15cm]{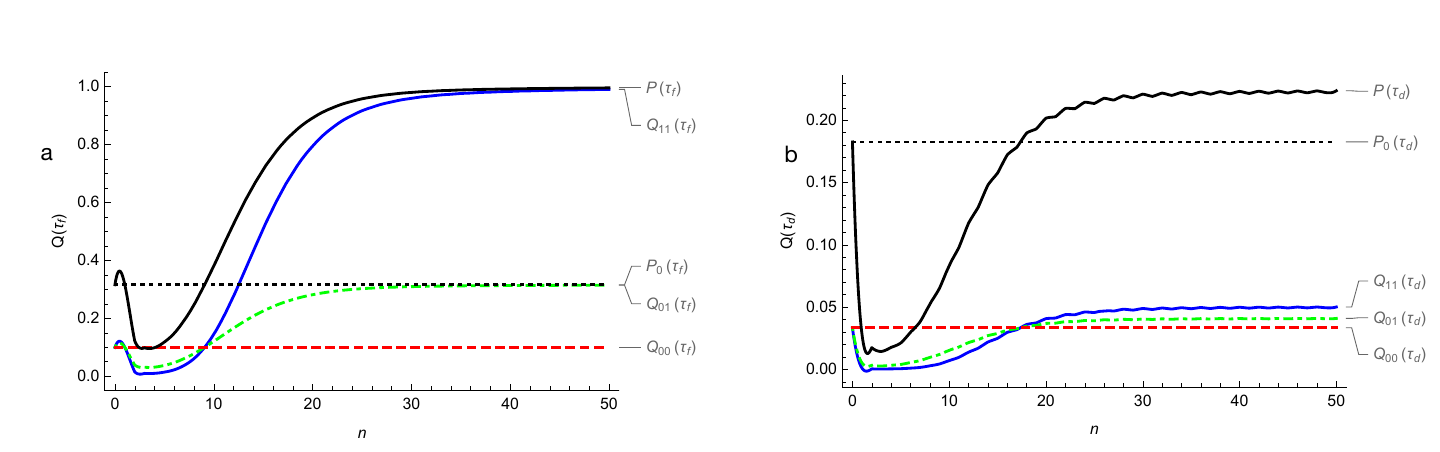}
\caption{ (Color online) Short-term regime (a) and long-term regime(b) behavior of the function $Q(t)$ (see eq. \ref{Qt}) with respect to applied pulse number, $n$, in the Markovian regime, for different scenarios, discussed in text.                                                                                                                                    The parameter $\eta$ set as $0.5$ and $t$ is dimensionless time in the unit of $\omega_c$.}
\label{fig1}
\end{figure}

\section{Results and Discussion}
In order to track the suppression of the undesired effects of dephasing by applying PDD, it is worth to study the behavior of the function
\begin{equation}\label{Qt}
Q(t)=
\begin{cases}
P(t) & \text {for single qubit system,}\\
P^1(t)P^2(t) & \text{for two-qubit system,}
\end{cases}
\end{equation}

\begin{figure}
\centering
\includegraphics[width=15cm]{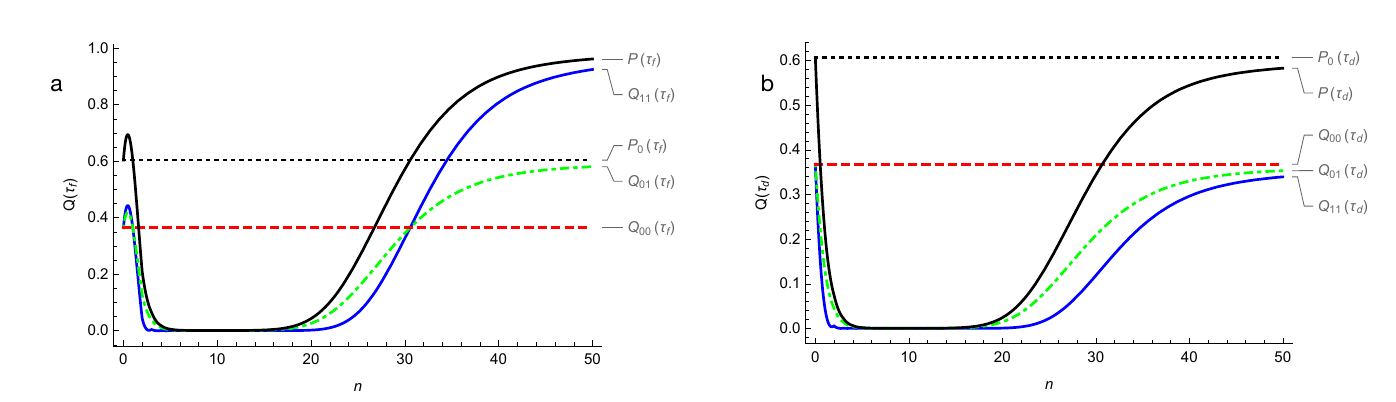}
\caption{ (Color online) Short-term regime (a) and long-term regime(b) behavior of the function $Q(t)$ (see eq. \ref{Qt}) with respect to applied pulse number, $n$, in the non-Markovian regime, for different scenarios, discussed in text.                                                                                                                                    The parameter $\eta$ set as $0.5$ and $t$ is dimensionless time in the unit of $\omega_c$.}
\label{fig2}
\end{figure}

for four different possible ways to applying PDD pulses on qubits: \textit{i)} absence of PDD pulses ($Q_{00}$ protocol),\textit{ ii)} applying PDD pulses only on first qubit ($Q_{10}$ protocol), \textit{iii)} applying PDD pulses only on second qubit ($Q_{01}$ protocol) and \textit{iv)} applying PDD pulses on both qubits ($Q_{11}$ protocol). Symmetry implies that $Q_{01}\equiv Q_{10}$. Figures (\ref{fig1}) and (\ref{fig2}) compare the behavior of $Q(t)$ for these scenarios in the both short and long term regimes and for both Markovian and non-Markovian cases. The results are in agreement with the previous works for the single-qubit case\cite{song2017control}. For the case of two-qubit systems, the results reveal that increasing the number of pulses on both qubits removes the dephasing effects completely, i.e. $Q(t)\longrightarrow 1$ as $n$ increases. But if the pulses applied only on one qubit we can control only the fifty percent of dephasing process, under the best conditions. Furthermore, like the single-qubit case, for small value of pulse number, $Q(t)$ vanishes and quantum fluctuation ceases \textit{i.e.} for the regime of few number of pulses, PDD accomplices with decoherence to kill the quantum coherence of the system. But it could suppress the decoherence for large number of pulses. Furthermore, the number of pulses required for decoherence suppression is more for non-Markovian dynamics.

\begin{figure}
\centering
\includegraphics[width=15cm]{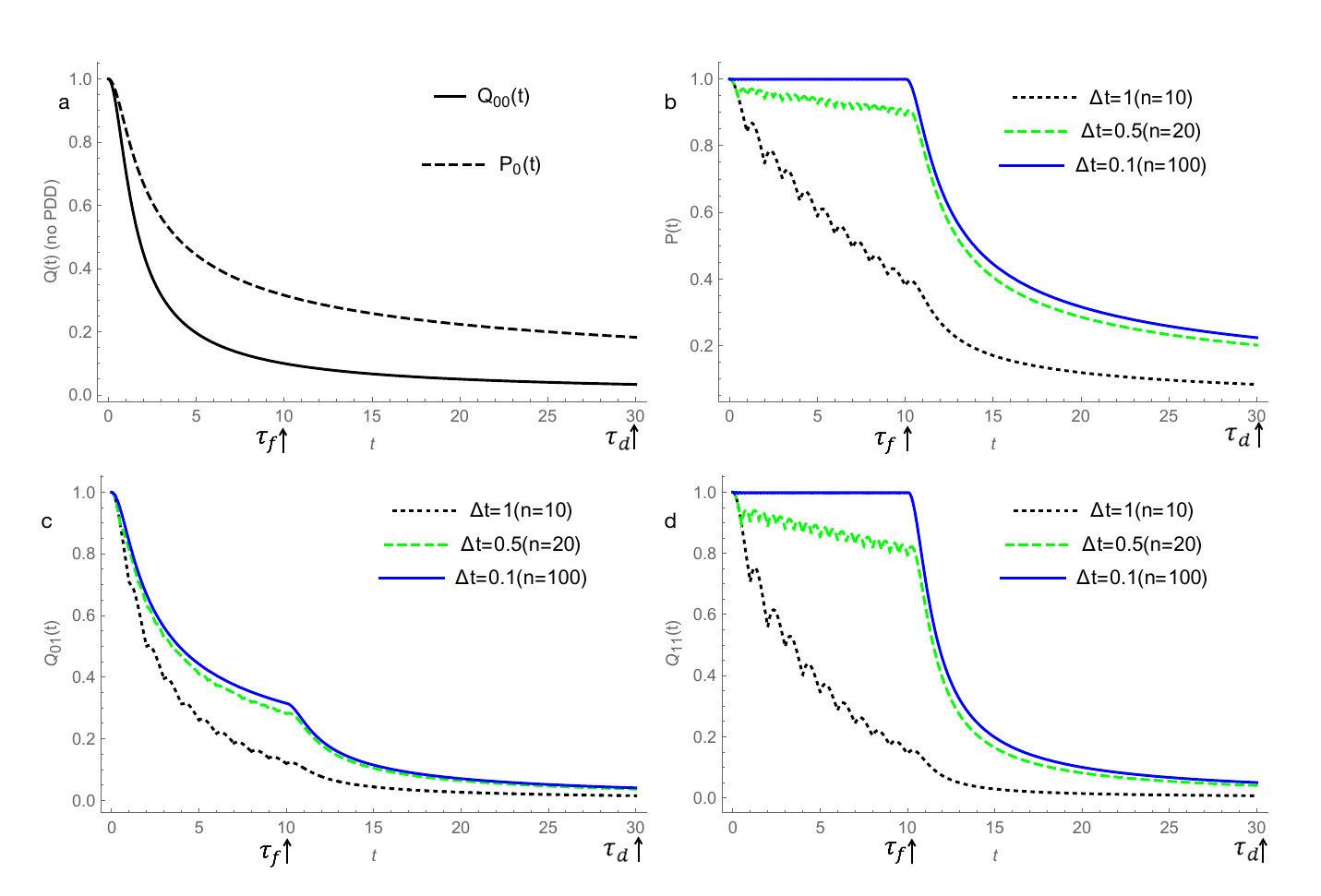}
\caption{ (Color online) Dynamics of $Q(t)$ (see eq. \ref{Qt})in the Markovian regime ($s=1$) for various scenarios; when a) no PDD applied to the qubit(s) ($P_0(t)$ refers to single qubit case and $Q_{00}$ is for two-qubit system,) b)the PDD applied to single qubit system. In the two-qubit system c) the PDD applied only on one of the qubits ($Q_{01/10}$ protocol) d) the PDD applied to both qubits ($Q_{11}$ protocol) for different pulse separations: $\Delta t=1$ \textit{i.e.} 10 pulses are applied during time interval $ [0,\tau_f] $ (black-dotted line), $\Delta t=0.5$ \textit{i.e.} 20 pulses are applied during time interval $ [0,\tau_f] $ (green dashed line) and $\Delta t=0.1$ \textit{i.e.} 100 pulses are applied during time interval $ [0,\tau_f] $ (Blue line). The parameter $\eta$ set as $0.5$ and $t$ is dimensionless time in the unit of $\omega_c$. }
\label{fig3}
\end{figure}

\begin{figure}
\centering
\includegraphics[width=15cm]{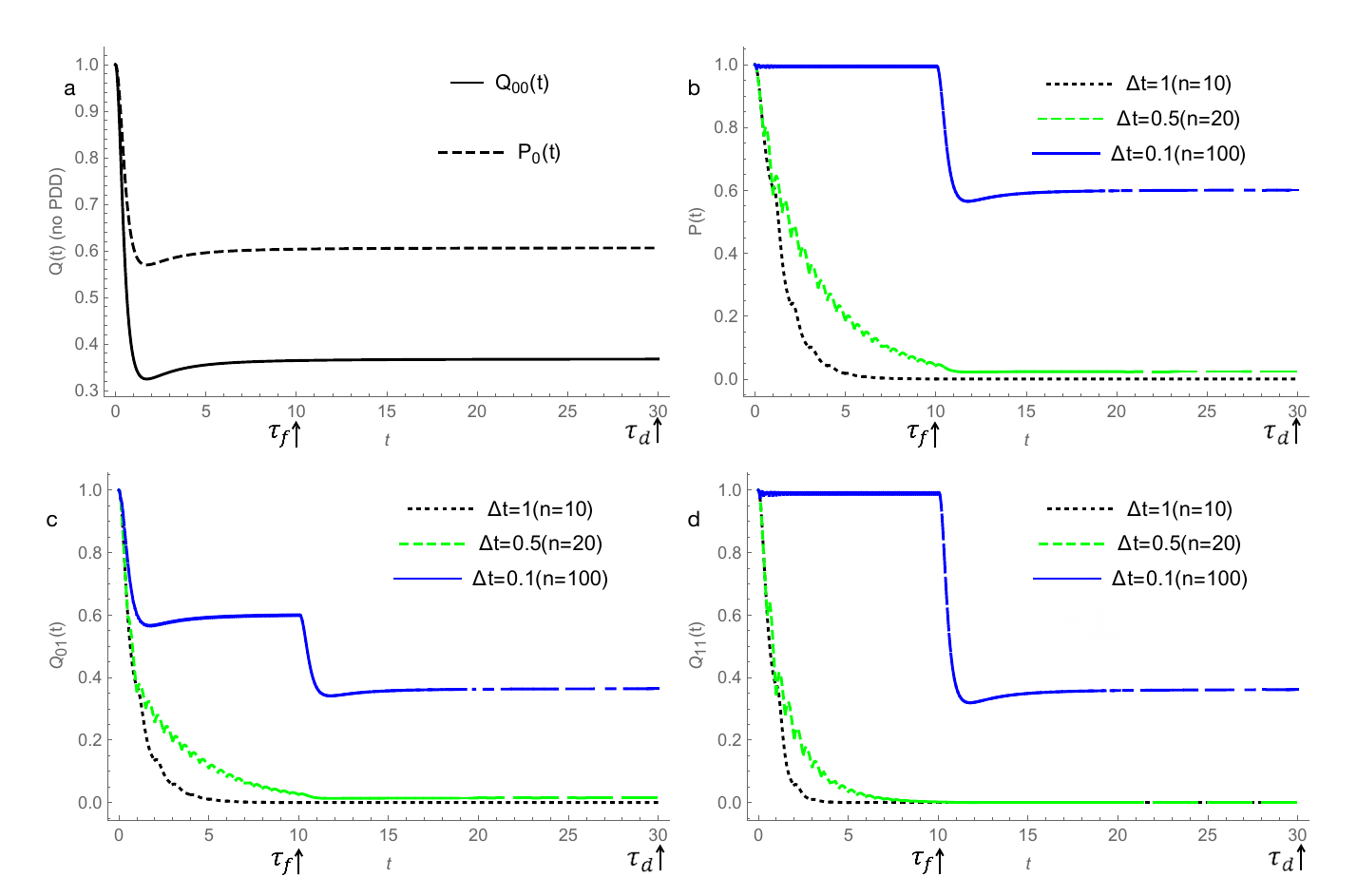}
\caption{ (Color online) Dynamics of $Q(t)$ (see eq. \ref{Qt})in the non-Markovian regime ($s=3$) for various scenarios; when a) no PDD applied to the qubit(s) ($P_0(t)$ refers to single qubit case and $Q_{00}$ is for two-qubit system,) b) the PDD applied to single qubit system. In the two-qubit system c) the PDD applied only on one of the qubits ($Q_{01/10}$ protocol) d) the PDD applied to both qubits ($Q_{11}$ protocol) for different pulse separations: $\Delta t=1$ \textit{i.e.} 10 pulses are applied during time interval $ [0,\tau_f] $ (black-dotted line), $\Delta t=0.5$ \textit{i.e.} 20 pulses are applied during time interval $ [0,\tau_f] $ (green dashed line) and $\Delta t=0.1$ \textit{i.e.} 100 pulses are applied during time interval $ [0,\tau_f] $ (Blue line). The parameter $\eta$ set as $0.5$ and $t$ is dimensionless time in the unit of $\omega_c$.}
\label{fig4}
\end{figure}

The effects of duty cycle (or repetition rate) of applied pulses are depicted in figures (\ref{fig3})and (\ref{fig4}). Each pulse tends to correct the error raised by dephasing process but during the time interval between pulses, $\Delta t=[\frac{N}{\tau_f}]$, leakage of the quantum coherence to the environment causes the dephasing and hence lost of quantum information encoded in the qubit(s) system. Increasing the pulse repetition rate (decreasing $\Delta t$) causes the information leakage to stop and hence more system- environment decouplig occurs. Since the number of applied pulses is essential for maintaining the quantum coherence of the system, higher pulse repetition rate (i.e. smaller values of $\Delta t$) causes to reach the enough number of the PDD pulses in shorter time and hence improves the coherence preservation process due to PDD procedure.
The results show that for short-term regime, where the train of PDDs continues until the observation time, \textit{i.e} $\tau_d=\tau_f$ and for  enough number of decoupling pulses, PDD process cancels the pure dephasing effects, i.e. the PDD process recovers/freezes the system into/on its initial state. Hence, all quantum features of the system recovered/preserved during the evolution. In the long-term dynamical regime, the state of the system and its quantum features evolves under dephasing process induced by slightly modified environment(s), after stopping the pulses. In this regime, the dephasing rate is modified to  lower values with respect to intact system.

\begin{figure}
\centering
\includegraphics[width=17cm]{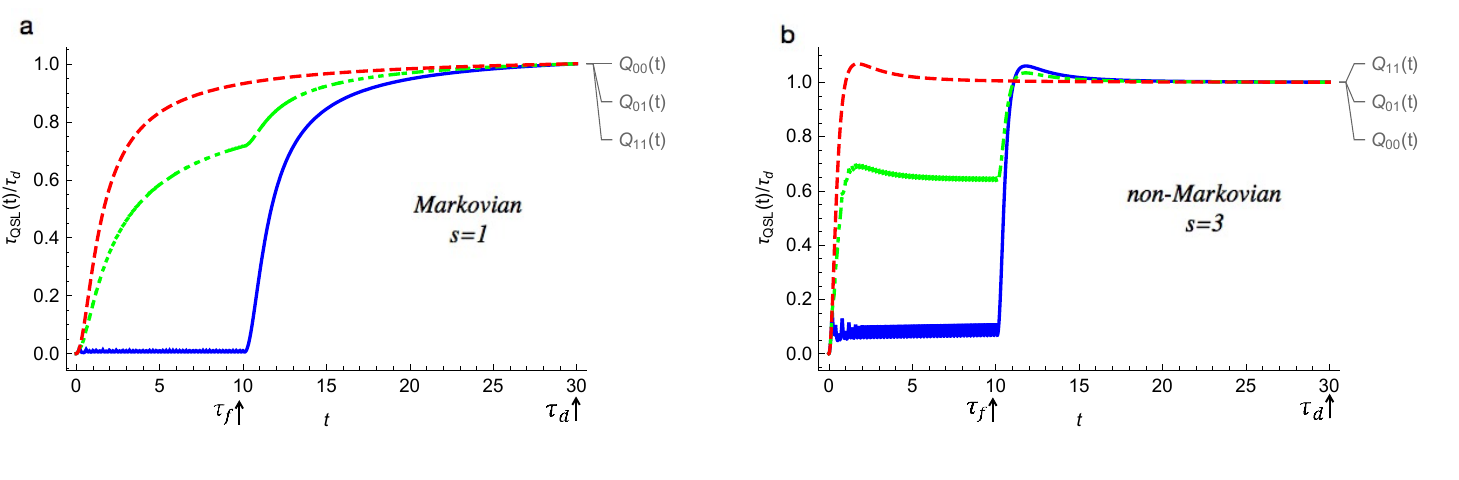}
\caption{ (Color online)Time evolution of the normalized QSLT upper bound for both Markovian regime (a) and non-Markovian regime (b), for different scenarios, discussed in text. The PDD pulses are applied during time interval $ [0,\tau_f]$ on the system which is prepared in singlet state, initially. The parameter $\eta$ set as $0.5$ and $t$ is dimensionless time in the unit of $\omega_c$. }
\label{fig5}
\end{figure}

\begin{figure}
\centering
\includegraphics[width=17cm]{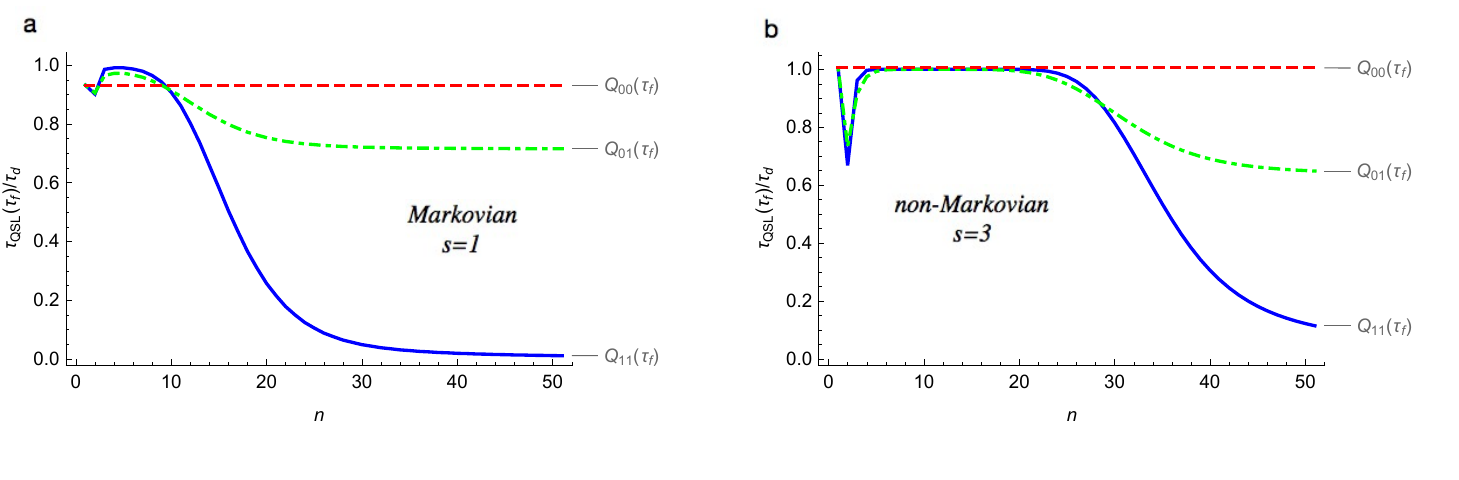}
\caption{ (Color online) Normalized QSLT upper bound vs. pulse number, n,for both Markovian regime (a) and non-Markovian regime (b), for different scenarios, discussed in text. The PDD pulses are applied during time interval $ [0,\tau_f]$ on the system which is prepared in singlet state, initially. The parameter $\eta$ set as $0.5$ and $t$ is dimensionless time in the unit of $\omega_c$.}
\label{fig6}
\end{figure}

The effect of DD pulse on the QSLT of the system is also depicted in figure (\ref{fig5}) for both Markovian and non-Markovian dynamical regimes. This figure shows that when the pulses are applied to both qubits ($Q_{11}$ protocol) the QSLT approaches to zero during the PDD process (short term regime ($t<\tau_f=10$)). This means that the quantum system can evolve approximately in instant times. In other word, The QSL approaches to infinity. This does not conflict with the Heisenberg uncertainty relation, because during the PDD process the energy of the system becomes undefined and hence $\Delta E \longrightarrow \infty$. Additionally, here we consider only the dephasing process due to system-environment interaction, which can be surpassed by decoupling process, and does not consider intrinsic decoherence, which implies the existence of a minimum time scale for the time evolution of quantum systems (coarse grained dynamics)\cite{EPJD2011,PhysRevA.44.5401}.  In long term regime ($\tau_f=10 \omega_c< time < \tau_d=30 \omega_c$) the value of QSLT increases with time but still is smaller than $\tau_d$, which means the system has potential to speed up. Figure (\ref{fig6}) reveals that this speed up can be reached when an enough number of pulses are applied. This acceleration can be achieved for larger number of DD pulses for non-Markovian case because of the information bounce and back process between the system and surrounding environment. These figures, also, reveal that the $Q_{01/10}$ scheme is still useful to speedup the quantum computation and information processing tasks.

\section{Conclusion}
The influence of the periodic pulse decoupling (PDD) on the dynamics of the quantum correlation and the quantum speed limit time(QSLT) are investigated. The system includes two non-interacting qubits, each of them are embedded in its own bosonic bath. Pure dephasing process establishes the flux of the quantum information from the quantum system to the surrounding bath via Markovian/non-Markovian mechanisms. The solution is obtained by solving governing master equation or Kraus operator method. The results reveal that for short-term regime, where the train of PDDs continues until the observation time and enough number of decoupling pulses, PDDs cancel the pure dephasing effects, i.e. the PDD process recovers/freezes the system into/on its initial state. Hence, all quantum features of the system recovered/preserved during the evolution. In the long-term dynamical regime, the state of the system and its quantum features evolves under dephasing process induced by slightly modified environment(s), after stoping the PDDs. Also, It is possible to achieve ultra-high acceleration of quantum evolution.  
\nocite{*}
\section*{Contribution}
The present study was a collaborative effort among the authors, who all contributed to the preparation of the manuscript. The idea was proposed by H. M, while both H.M and A.A carried out the calculation and presented the data. H. M subsequently explained the results, and wrote the paper. The authors confirm that they have read and approved the final version of the manuscript.
\section*{Funding}
This research received no specific grant from any funding agency in the public, commercial, or not-for-profit sectors.

\section*{Disclosures} 
The authors declare no conﬂicts of interest.

\section*{Data availability} 
Data underlying the results presented in this paper are not publicly available but data is provided by corresponding author via email (h.r.mhmdi@gmail.com)

\bibliography{Mybib.bib}

\begin{thebibliography}{51}%
\makeatletter
\providecommand \@ifxundefined [1]{%
 \@ifx{#1\undefined}
}%
\providecommand \@ifnum [1]{%
 \ifnum #1\expandafter \@firstoftwo
 \else \expandafter \@secondoftwo
 \fi
}%
\providecommand \@ifx [1]{%
 \ifx #1\expandafter \@firstoftwo
 \else \expandafter \@secondoftwo
 \fi
}%
\providecommand \natexlab [1]{#1}%
\providecommand \enquote  [1]{``#1''}%
\providecommand \bibnamefont  [1]{#1}%
\providecommand \bibfnamefont [1]{#1}%
\providecommand \citenamefont [1]{#1}%
\providecommand \href@noop [0]{\@secondoftwo}%
\providecommand \href [0]{\begingroup \@sanitize@url \@href}%
\providecommand \@href[1]{\@@startlink{#1}\@@href}%
\providecommand \@@href[1]{\endgroup#1\@@endlink}%
\providecommand \@sanitize@url [0]{\catcode `\\12\catcode `\$12\catcode
  `\&12\catcode `\#12\catcode `\^12\catcode `\_12\catcode `\%12\relax}%
\providecommand \@@startlink[1]{}%
\providecommand \@@endlink[0]{}%
\providecommand \url  [0]{\begingroup\@sanitize@url \@url }%
\providecommand \@url [1]{\endgroup\@href {#1}{\urlprefix }}%
\providecommand \urlprefix  [0]{URL }%
\providecommand \Eprint [0]{\href }%
\providecommand \doibase [0]{https://doi.org/}%
\providecommand \selectlanguage [0]{\@gobble}%
\providecommand \bibinfo  [0]{\@secondoftwo}%
\providecommand \bibfield  [0]{\@secondoftwo}%
\providecommand \translation [1]{[#1]}%
\providecommand \BibitemOpen [0]{}%
\providecommand \bibitemStop [0]{}%
\providecommand \bibitemNoStop [0]{.\EOS\space}%
\providecommand \EOS [0]{\spacefactor3000\relax}%
\providecommand \BibitemShut  [1]{\csname bibitem#1\endcsname}%
\let\auto@bib@innerbib\@empty
%</preamble>
\bibitem [{\citenamefont {Zhang}\ \emph {et~al.}(2014)\citenamefont {Zhang},
  \citenamefont {Han}, \citenamefont {Xia}, \citenamefont {Cao},\ and\
  \citenamefont {Fan}}]{zhang.srep2014}%
  \BibitemOpen
  \bibfield  {author} {\bibinfo {author} {\bibfnamefont {Y.-J.}\ \bibnamefont
  {Zhang}}, \bibinfo {author} {\bibfnamefont {W.}~\bibnamefont {Han}}, \bibinfo
  {author} {\bibfnamefont {Y.-J.}\ \bibnamefont {Xia}}, \bibinfo {author}
  {\bibfnamefont {J.-P.}\ \bibnamefont {Cao}},\ and\ \bibinfo {author}
  {\bibfnamefont {H.}~\bibnamefont {Fan}},\ }\bibfield  {title} {\bibinfo
  {title} {Quantum speed limit for arbitrary initial states},\ }\href@noop {}
  {\bibfield  {journal} {\bibinfo  {journal} {Scientific Reports}\ }\textbf
  {\bibinfo {volume} {4}},\ \bibinfo {pages} {1} (\bibinfo {year}
  {2014})}\BibitemShut {NoStop}%
\bibitem [{\citenamefont
  {Bekenstein}(1981)}]{zhang.srep1-bekenstein1981energy}%
  \BibitemOpen
  \bibfield  {author} {\bibinfo {author} {\bibfnamefont {J.~D.}\ \bibnamefont
  {Bekenstein}},\ }\bibfield  {title} {\bibinfo {title} {Energy cost of
  information transfer},\ }\href@noop {} {\bibfield  {journal} {\bibinfo
  {journal} {Physical Review Letters}\ }\textbf {\bibinfo {volume} {46}},\
  \bibinfo {pages} {623} (\bibinfo {year} {1981})}\BibitemShut {NoStop}%
\bibitem [{\citenamefont {Giovannetti}\ \emph {et~al.}(2011)\citenamefont
  {Giovannetti}, \citenamefont {Lloyd},\ and\ \citenamefont
  {Maccone}}]{zhang.srep2-giovannetti2011advances}%
  \BibitemOpen
  \bibfield  {author} {\bibinfo {author} {\bibfnamefont {V.}~\bibnamefont
  {Giovannetti}}, \bibinfo {author} {\bibfnamefont {S.}~\bibnamefont {Lloyd}},\
  and\ \bibinfo {author} {\bibfnamefont {L.}~\bibnamefont {Maccone}},\
  }\bibfield  {title} {\bibinfo {title} {Advances in quantum metrology},\
  }\href@noop {} {\bibfield  {journal} {\bibinfo  {journal} {Nature Photonics}\
  }\textbf {\bibinfo {volume} {5}},\ \bibinfo {pages} {222} (\bibinfo {year}
  {2011})}\BibitemShut {NoStop}%
\bibitem [{\citenamefont {Lloyd}(2002)}]{zhang.srep3-lloyd2002computational}%
  \BibitemOpen
  \bibfield  {author} {\bibinfo {author} {\bibfnamefont {S.}~\bibnamefont
  {Lloyd}},\ }\bibfield  {title} {\bibinfo {title} {Computational capacity of
  the universe},\ }\href@noop {} {\bibfield  {journal} {\bibinfo  {journal}
  {Physical Review Letters}\ }\textbf {\bibinfo {volume} {88}},\ \bibinfo
  {pages} {237901} (\bibinfo {year} {2002})}\BibitemShut {NoStop}%
\bibitem [{\citenamefont {Caneva}\ \emph {et~al.}(2009)\citenamefont {Caneva},
  \citenamefont {Murphy}, \citenamefont {Calarco}, \citenamefont {Fazio},
  \citenamefont {Montangero}, \citenamefont {Giovannetti},\ and\ \citenamefont
  {Santoro}}]{zhang.srep4-caneva2009optimal}%
  \BibitemOpen
  \bibfield  {author} {\bibinfo {author} {\bibfnamefont {T.}~\bibnamefont
  {Caneva}}, \bibinfo {author} {\bibfnamefont {M.}~\bibnamefont {Murphy}},
  \bibinfo {author} {\bibfnamefont {T.}~\bibnamefont {Calarco}}, \bibinfo
  {author} {\bibfnamefont {R.}~\bibnamefont {Fazio}}, \bibinfo {author}
  {\bibfnamefont {S.}~\bibnamefont {Montangero}}, \bibinfo {author}
  {\bibfnamefont {V.}~\bibnamefont {Giovannetti}},\ and\ \bibinfo {author}
  {\bibfnamefont {G.~E.}\ \bibnamefont {Santoro}},\ }\bibfield  {title}
  {\bibinfo {title} {Optimal control at the quantum speed limit},\ }\href@noop
  {} {\bibfield  {journal} {\bibinfo  {journal} {Physical Review Letters}\
  }\textbf {\bibinfo {volume} {103}},\ \bibinfo {pages} {240501} (\bibinfo
  {year} {2009})}\BibitemShut {NoStop}%
\bibitem [{\citenamefont {Mandelstam}\ and\ \citenamefont
  {Tamm}(1945)}]{zhang.srep5-mandelstam1945energy}%
  \BibitemOpen
  \bibfield  {author} {\bibinfo {author} {\bibfnamefont {L.}~\bibnamefont
  {Mandelstam}}\ and\ \bibinfo {author} {\bibfnamefont {I.}~\bibnamefont
  {Tamm}},\ }\bibfield  {title} {\bibinfo {title} {The energy--time uncertainty
  relation in non-relativistic quantum mechanics},\ }\href@noop {} {\bibfield
  {journal} {\bibinfo  {journal} {Izv. Akad. Nauk SSSR}\ }\textbf {\bibinfo
  {volume} {9}},\ \bibinfo {pages} {122} (\bibinfo {year} {1945})}\BibitemShut
  {NoStop}%
\bibitem [{\citenamefont {Fleming}(1973)}]{zhang.srep6-fleming1973unitarity}%
  \BibitemOpen
  \bibfield  {author} {\bibinfo {author} {\bibfnamefont {G.~N.}\ \bibnamefont
  {Fleming}},\ }\bibfield  {title} {\bibinfo {title} {A unitarity bound on the
  evolution of nonstationary states},\ }\href@noop {} {\bibfield  {journal}
  {\bibinfo  {journal} {Il Nuovo Cimento A (1965-1970)}\ }\textbf {\bibinfo
  {volume} {16}},\ \bibinfo {pages} {232} (\bibinfo {year} {1973})}\BibitemShut
  {NoStop}%
\bibitem [{\citenamefont {Anandan}\ and\ \citenamefont
  {Aharonov}(1990)}]{zhang.srep7-anandan1990geometry}%
  \BibitemOpen
  \bibfield  {author} {\bibinfo {author} {\bibfnamefont {J.}~\bibnamefont
  {Anandan}}\ and\ \bibinfo {author} {\bibfnamefont {Y.}~\bibnamefont
  {Aharonov}},\ }\bibfield  {title} {\bibinfo {title} {Geometry of quantum
  evolution},\ }\href@noop {} {\bibfield  {journal} {\bibinfo  {journal}
  {Physical Review Letters}\ }\textbf {\bibinfo {volume} {65}},\ \bibinfo
  {pages} {1697} (\bibinfo {year} {1990})}\BibitemShut {NoStop}%
\bibitem [{\citenamefont {Vaidman}(1992)}]{zhang.srep8-vaidman1992minimum}%
  \BibitemOpen
  \bibfield  {author} {\bibinfo {author} {\bibfnamefont {L.}~\bibnamefont
  {Vaidman}},\ }\bibfield  {title} {\bibinfo {title} {Minimum time for the
  evolution to an orthogonal quantum state},\ }\href@noop {} {\bibfield
  {journal} {\bibinfo  {journal} {American Journal of Physics}\ }\textbf
  {\bibinfo {volume} {60}},\ \bibinfo {pages} {182} (\bibinfo {year}
  {1992})}\BibitemShut {NoStop}%
\bibitem [{\citenamefont {Margolus}\ and\ \citenamefont
  {Levitin}(1998)}]{zhang.srep9-margolus1998maximum}%
  \BibitemOpen
  \bibfield  {author} {\bibinfo {author} {\bibfnamefont {N.}~\bibnamefont
  {Margolus}}\ and\ \bibinfo {author} {\bibfnamefont {L.~B.}\ \bibnamefont
  {Levitin}},\ }\bibfield  {title} {\bibinfo {title} {The maximum speed of
  dynamical evolution},\ }\href@noop {} {\bibfield  {journal} {\bibinfo
  {journal} {Physica D: Nonlinear Phenomena}\ }\textbf {\bibinfo {volume}
  {120}},\ \bibinfo {pages} {188} (\bibinfo {year} {1998})}\BibitemShut
  {NoStop}%
\bibitem [{\citenamefont {Levitin}\ and\ \citenamefont
  {Toffoli}(2009)}]{zhang.srep10-levitin2009fundamental}%
  \BibitemOpen
  \bibfield  {author} {\bibinfo {author} {\bibfnamefont {L.~B.}\ \bibnamefont
  {Levitin}}\ and\ \bibinfo {author} {\bibfnamefont {T.}~\bibnamefont
  {Toffoli}},\ }\bibfield  {title} {\bibinfo {title} {Fundamental limit on the
  rate of quantum dynamics: the unified bound is tight},\ }\href@noop {}
  {\bibfield  {journal} {\bibinfo  {journal} {Physical Review Letters}\
  }\textbf {\bibinfo {volume} {103}},\ \bibinfo {pages} {160502} (\bibinfo
  {year} {2009})}\BibitemShut {NoStop}%
\bibitem [{\citenamefont {Giovannetti}\ \emph {et~al.}(2003)\citenamefont
  {Giovannetti}, \citenamefont {Lloyd},\ and\ \citenamefont
  {Maccone}}]{zhang.srep11-PhysRevA.67.052109}%
  \BibitemOpen
  \bibfield  {author} {\bibinfo {author} {\bibfnamefont {V.}~\bibnamefont
  {Giovannetti}}, \bibinfo {author} {\bibfnamefont {S.}~\bibnamefont {Lloyd}},\
  and\ \bibinfo {author} {\bibfnamefont {L.}~\bibnamefont {Maccone}},\
  }\bibfield  {title} {\bibinfo {title} {Quantum limits to dynamical
  evolution},\ }\href {https://doi.org/10.1103/PhysRevA.67.052109} {\bibfield
  {journal} {\bibinfo  {journal} {Physical Review A}\ }\textbf {\bibinfo
  {volume} {67}},\ \bibinfo {pages} {052109} (\bibinfo {year}
  {2003})}\BibitemShut {NoStop}%
\bibitem [{\citenamefont {Jones}\ and\ \citenamefont
  {Kok}(2010)}]{zhang.srep12-PhysRevA.82.022107}%
  \BibitemOpen
  \bibfield  {author} {\bibinfo {author} {\bibfnamefont {P.~J.}\ \bibnamefont
  {Jones}}\ and\ \bibinfo {author} {\bibfnamefont {P.}~\bibnamefont {Kok}},\
  }\bibfield  {title} {\bibinfo {title} {Geometric derivation of the quantum
  speed limit},\ }\href {https://doi.org/10.1103/PhysRevA.82.022107} {\bibfield
   {journal} {\bibinfo  {journal} {Physical Review A}\ }\textbf {\bibinfo
  {volume} {82}},\ \bibinfo {pages} {022107} (\bibinfo {year}
  {2010})}\BibitemShut {NoStop}%
\bibitem [{\citenamefont {Deffner}\ and\ \citenamefont
  {Lutz}(2013{\natexlab{a}})}]{zhang.srep13-Deffner_2013}%
  \BibitemOpen
  \bibfield  {author} {\bibinfo {author} {\bibfnamefont {S.}~\bibnamefont
  {Deffner}}\ and\ \bibinfo {author} {\bibfnamefont {E.}~\bibnamefont {Lutz}},\
  }\bibfield  {title} {\bibinfo {title} {Energy–time uncertainty relation for
  driven quantum systems},\ }\href
  {https://doi.org/10.1088/1751-8113/46/33/335302} {\bibfield  {journal}
  {\bibinfo  {journal} {Journal of Physics A: Mathematical and Theoretical}\
  }\textbf {\bibinfo {volume} {46}},\ \bibinfo {pages} {335302} (\bibinfo
  {year} {2013}{\natexlab{a}})}\BibitemShut {NoStop}%
\bibitem [{\citenamefont {Pfeifer}(1993)}]{zhang.srep14-PhysRevLett.70.3365}%
  \BibitemOpen
  \bibfield  {author} {\bibinfo {author} {\bibfnamefont {P.}~\bibnamefont
  {Pfeifer}},\ }\bibfield  {title} {\bibinfo {title} {How fast can a quantum
  state change with time?},\ }\href
  {https://doi.org/10.1103/PhysRevLett.70.3365} {\bibfield  {journal} {\bibinfo
   {journal} {Physical Review Letters}\ }\textbf {\bibinfo {volume} {70}},\
  \bibinfo {pages} {3365} (\bibinfo {year} {1993})}\BibitemShut {NoStop}%
\bibitem [{\citenamefont {Pfeifer}\ and\ \citenamefont
  {Fr\"ohlich}(1995)}]{zhang.srep15-RevModPhys.67.759}%
  \BibitemOpen
  \bibfield  {author} {\bibinfo {author} {\bibfnamefont {P.}~\bibnamefont
  {Pfeifer}}\ and\ \bibinfo {author} {\bibfnamefont {J.}~\bibnamefont
  {Fr\"ohlich}},\ }\bibfield  {title} {\bibinfo {title} {Generalized
  time-energy uncertainty relations and bounds on lifetimes of resonances},\
  }\href {https://doi.org/10.1103/RevModPhys.67.759} {\bibfield  {journal}
  {\bibinfo  {journal} {Review Modern Physics}\ }\textbf {\bibinfo {volume}
  {67}},\ \bibinfo {pages} {759} (\bibinfo {year} {1995})}\BibitemShut
  {NoStop}%
\bibitem [{\citenamefont {Taddei}\ \emph {et~al.}(2013)\citenamefont {Taddei},
  \citenamefont {Escher}, \citenamefont {Davidovich},\ and\ \citenamefont
  {de~Matos~Filho}}]{zhang.srep16-taddei2013quantum}%
  \BibitemOpen
  \bibfield  {author} {\bibinfo {author} {\bibfnamefont {M.~M.}\ \bibnamefont
  {Taddei}}, \bibinfo {author} {\bibfnamefont {B.~M.}\ \bibnamefont {Escher}},
  \bibinfo {author} {\bibfnamefont {L.}~\bibnamefont {Davidovich}},\ and\
  \bibinfo {author} {\bibfnamefont {R.~L.}\ \bibnamefont {de~Matos~Filho}},\
  }\bibfield  {title} {\bibinfo {title} {Quantum speed limit for physical
  processes},\ }\href@noop {} {\bibfield  {journal} {\bibinfo  {journal}
  {Physical Review Letters}\ }\textbf {\bibinfo {volume} {110}},\ \bibinfo
  {pages} {050402} (\bibinfo {year} {2013})}\BibitemShut {NoStop}%
\bibitem [{\citenamefont {del Campo}\ \emph {et~al.}(2013)\citenamefont {del
  Campo}, \citenamefont {Egusquiza}, \citenamefont {Plenio},\ and\
  \citenamefont {Huelga}}]{zhang.srep17-del2013quantum}%
  \BibitemOpen
  \bibfield  {author} {\bibinfo {author} {\bibfnamefont {A.}~\bibnamefont {del
  Campo}}, \bibinfo {author} {\bibfnamefont {I.~L.}\ \bibnamefont {Egusquiza}},
  \bibinfo {author} {\bibfnamefont {M.~B.}\ \bibnamefont {Plenio}},\ and\
  \bibinfo {author} {\bibfnamefont {S.~F.}\ \bibnamefont {Huelga}},\ }\bibfield
   {title} {\bibinfo {title} {Quantum speed limits in open system dynamics},\
  }\href@noop {} {\bibfield  {journal} {\bibinfo  {journal} {Physical Review
  Letters}\ }\textbf {\bibinfo {volume} {110}},\ \bibinfo {pages} {050403}
  (\bibinfo {year} {2013})}\BibitemShut {NoStop}%
\bibitem [{\citenamefont {Deffner}\ and\ \citenamefont
  {Lutz}(2013{\natexlab{b}})}]{zhang.srep18-deffner2013quantum}%
  \BibitemOpen
  \bibfield  {author} {\bibinfo {author} {\bibfnamefont {S.}~\bibnamefont
  {Deffner}}\ and\ \bibinfo {author} {\bibfnamefont {E.}~\bibnamefont {Lutz}},\
  }\bibfield  {title} {\bibinfo {title} {Quantum speed limit for non-markovian
  dynamics},\ }\href@noop {} {\bibfield  {journal} {\bibinfo  {journal}
  {Physical Review Letters}\ }\textbf {\bibinfo {volume} {111}},\ \bibinfo
  {pages} {010402} (\bibinfo {year} {2013}{\natexlab{b}})}\BibitemShut
  {NoStop}%
\bibitem [{\citenamefont {Schulman}(2008)}]{campo4-schulman2008jump}%
  \BibitemOpen
  \bibfield  {author} {\bibinfo {author} {\bibfnamefont {L.~S.}\ \bibnamefont
  {Schulman}},\ }\bibfield  {title} {\bibinfo {title} {Jump time and passage
  time: The duration ofs a quantum transition},\ }in\ \href@noop {} {\emph
  {\bibinfo {booktitle} {Time in Quantum Mechanics}}}\ (\bibinfo  {publisher}
  {Springer},\ \bibinfo {year} {2008})\ pp.\ \bibinfo {pages}
  {107--128}\BibitemShut {NoStop}%
\bibitem [{\citenamefont {Zwierz}(2012)}]{campo13-zwierz2012comment}%
  \BibitemOpen
  \bibfield  {author} {\bibinfo {author} {\bibfnamefont {M.}~\bibnamefont
  {Zwierz}},\ }\bibfield  {title} {\bibinfo {title} {Comment on “geometric
  derivation of the quantum speed limit”},\ }\href@noop {} {\bibfield
  {journal} {\bibinfo  {journal} {Physical Review A}\ }\textbf {\bibinfo
  {volume} {86}},\ \bibinfo {pages} {016101} (\bibinfo {year}
  {2012})}\BibitemShut {NoStop}%
\bibitem [{\citenamefont {Viola}\ and\ \citenamefont
  {Lloyd}(1998{\natexlab{a}})}]{addis1-viola1998}%
  \BibitemOpen
  \bibfield  {author} {\bibinfo {author} {\bibfnamefont {L.}~\bibnamefont
  {Viola}}\ and\ \bibinfo {author} {\bibfnamefont {S.}~\bibnamefont {Lloyd}},\
  }\bibfield  {title} {\bibinfo {title} {Dynamical suppression of decoherence
  in two-state quantum systems},\ }\href@noop {} {\bibfield  {journal}
  {\bibinfo  {journal} {Physical Review A}\ }\textbf {\bibinfo {volume} {58}},\
  \bibinfo {pages} {2733} (\bibinfo {year} {1998}{\natexlab{a}})}\BibitemShut
  {NoStop}%
\bibitem [{\citenamefont {Viola}\ \emph {et~al.}(1999)\citenamefont {Viola},
  \citenamefont {Knill},\ and\ \citenamefont {Lloyd}}]{addis2-viola1999}%
  \BibitemOpen
  \bibfield  {author} {\bibinfo {author} {\bibfnamefont {L.}~\bibnamefont
  {Viola}}, \bibinfo {author} {\bibfnamefont {E.}~\bibnamefont {Knill}},\ and\
  \bibinfo {author} {\bibfnamefont {S.}~\bibnamefont {Lloyd}},\ }\bibfield
  {title} {\bibinfo {title} {Dynamical decoupling of open quantum systems},\
  }\href@noop {} {\bibfield  {journal} {\bibinfo  {journal} {Physical Review
  Letters}\ }\textbf {\bibinfo {volume} {82}},\ \bibinfo {pages} {2417}
  (\bibinfo {year} {1999})}\BibitemShut {NoStop}%
\bibitem [{\citenamefont {Yin}\ \emph {et~al.}(2019)\citenamefont {Yin},
  \citenamefont {Song},\ and\ \citenamefont {Liu}}]{YIN2019136}%
  \BibitemOpen
  \bibfield  {author} {\bibinfo {author} {\bibfnamefont {S.}~\bibnamefont
  {Yin}}, \bibinfo {author} {\bibfnamefont {J.}~\bibnamefont {Song}},\ and\
  \bibinfo {author} {\bibfnamefont {S.}~\bibnamefont {Liu}},\ }\bibfield
  {title} {\bibinfo {title} {Quantum criticality of quantum speed limit for a
  two-qubit system in the spin chain with the dzyaloshinsky–moriya
  interaction},\ }\href
  {https://doi.org/https://doi.org/10.1016/j.physleta.2018.10.027} {\bibfield
  {journal} {\bibinfo  {journal} {Physics Letters A}\ }\textbf {\bibinfo
  {volume} {383}},\ \bibinfo {pages} {136} (\bibinfo {year}
  {2019})}\BibitemShut {NoStop}%
\bibitem [{\citenamefont {Howard}\ \emph {et~al.}(2023)\citenamefont {Howard},
  \citenamefont {Lidiak}, \citenamefont {Jameson}, \citenamefont {Basyildiz},
  \citenamefont {Clark}, \citenamefont {Zhao}, \citenamefont {Bal},
  \citenamefont {Long}, \citenamefont {Pappas}, \citenamefont {Singh},\ and\
  \citenamefont {Gong}}]{howard2023implementing}%
  \BibitemOpen
  \bibfield  {author} {\bibinfo {author} {\bibfnamefont {J.}~\bibnamefont
  {Howard}}, \bibinfo {author} {\bibfnamefont {A.}~\bibnamefont {Lidiak}},
  \bibinfo {author} {\bibfnamefont {C.}~\bibnamefont {Jameson}}, \bibinfo
  {author} {\bibfnamefont {B.}~\bibnamefont {Basyildiz}}, \bibinfo {author}
  {\bibfnamefont {K.}~\bibnamefont {Clark}}, \bibinfo {author} {\bibfnamefont
  {T.}~\bibnamefont {Zhao}}, \bibinfo {author} {\bibfnamefont {M.}~\bibnamefont
  {Bal}}, \bibinfo {author} {\bibfnamefont {J.}~\bibnamefont {Long}}, \bibinfo
  {author} {\bibfnamefont {D.~P.}\ \bibnamefont {Pappas}}, \bibinfo {author}
  {\bibfnamefont {M.}~\bibnamefont {Singh}},\ and\ \bibinfo {author}
  {\bibfnamefont {Z.}~\bibnamefont {Gong}},\ }\href@noop {} {\bibinfo {title}
  {Implementing two-qubit gates at the quantum speed limit}} (\bibinfo {year}
  {2023}),\ \Eprint {https://arxiv.org/abs/2206.07716} {arXiv:2206.07716
  [quant-ph]} \BibitemShut {NoStop}%
\bibitem [{\citenamefont {Barthel}\ \emph {et~al.}(2022)\citenamefont
  {Barthel}, \citenamefont {Huber}, \citenamefont {Casanova}, \citenamefont
  {Arrazola}, \citenamefont {Niroomand}, \citenamefont {Sriarunothai},
  \citenamefont {Plenio},\ and\ \citenamefont
  {Wunderlich}}]{barthel2022robust}%
  \BibitemOpen
  \bibfield  {author} {\bibinfo {author} {\bibfnamefont {P.}~\bibnamefont
  {Barthel}}, \bibinfo {author} {\bibfnamefont {P.~H.}\ \bibnamefont {Huber}},
  \bibinfo {author} {\bibfnamefont {J.}~\bibnamefont {Casanova}}, \bibinfo
  {author} {\bibfnamefont {I.}~\bibnamefont {Arrazola}}, \bibinfo {author}
  {\bibfnamefont {D.}~\bibnamefont {Niroomand}}, \bibinfo {author}
  {\bibfnamefont {T.}~\bibnamefont {Sriarunothai}}, \bibinfo {author}
  {\bibfnamefont {M.~B.}\ \bibnamefont {Plenio}},\ and\ \bibinfo {author}
  {\bibfnamefont {C.}~\bibnamefont {Wunderlich}},\ }\href@noop {} {\bibinfo
  {title} {Robust two-qubit gates using pulsed dynamical decoupling}} (\bibinfo
  {year} {2022}),\ \Eprint {https://arxiv.org/abs/2208.00187} {arXiv:2208.00187
  [quant-ph]} \BibitemShut {NoStop}%
\bibitem [{\citenamefont {Ashhab}\ \emph {et~al.}(2022)\citenamefont {Ashhab},
  \citenamefont {Yoshihara}, \citenamefont {Fuse}, \citenamefont {Yamamoto},
  \citenamefont {Lupascu},\ and\ \citenamefont {Semba}}]{PhysRevA.105.042614}%
  \BibitemOpen
  \bibfield  {author} {\bibinfo {author} {\bibfnamefont {S.}~\bibnamefont
  {Ashhab}}, \bibinfo {author} {\bibfnamefont {F.}~\bibnamefont {Yoshihara}},
  \bibinfo {author} {\bibfnamefont {T.}~\bibnamefont {Fuse}}, \bibinfo {author}
  {\bibfnamefont {N.}~\bibnamefont {Yamamoto}}, \bibinfo {author}
  {\bibfnamefont {A.}~\bibnamefont {Lupascu}},\ and\ \bibinfo {author}
  {\bibfnamefont {K.}~\bibnamefont {Semba}},\ }\bibfield  {title} {\bibinfo
  {title} {Speed limits for two-qubit gates with weakly anharmonic qubits},\
  }\href {https://doi.org/10.1103/PhysRevA.105.042614} {\bibfield  {journal}
  {\bibinfo  {journal} {Physical Review A}\ }\textbf {\bibinfo {volume}
  {105}},\ \bibinfo {pages} {042614} (\bibinfo {year} {2022})}\BibitemShut
  {NoStop}%
\bibitem [{\citenamefont {Hegde}\ \emph {et~al.}(2022)\citenamefont {Hegde},
  \citenamefont {Zhang},\ and\ \citenamefont {Suter}}]{PhysRevLett.128.230502}%
  \BibitemOpen
  \bibfield  {author} {\bibinfo {author} {\bibfnamefont {S.~S.}\ \bibnamefont
  {Hegde}}, \bibinfo {author} {\bibfnamefont {J.}~\bibnamefont {Zhang}},\ and\
  \bibinfo {author} {\bibfnamefont {D.}~\bibnamefont {Suter}},\ }\bibfield
  {title} {\bibinfo {title} {Toward the speed limit of high-fidelity two-qubit
  gates},\ }\href {https://doi.org/10.1103/PhysRevLett.128.230502} {\bibfield
  {journal} {\bibinfo  {journal} {Physical Review Letters}\ }\textbf {\bibinfo
  {volume} {128}},\ \bibinfo {pages} {230502} (\bibinfo {year}
  {2022})}\BibitemShut {NoStop}%
\bibitem [{\citenamefont {Slichter}(2013)}]{slichter2013}%
  \BibitemOpen
  \bibfield  {author} {\bibinfo {author} {\bibfnamefont {C.~P.}\ \bibnamefont
  {Slichter}},\ }\href@noop {} {\emph {\bibinfo {title} {Principles of magnetic
  resonance}}},\ Vol.~\bibinfo {volume} {1}\ (\bibinfo  {publisher} {Springer
  Science \& Business Media},\ \bibinfo {year} {2013})\BibitemShut {NoStop}%
\bibitem [{\citenamefont {Presilla}\ \emph {et~al.}(1996)\citenamefont
  {Presilla}, \citenamefont {Onofrio},\ and\ \citenamefont
  {Tambini}}]{viola20-presilla1996}%
  \BibitemOpen
  \bibfield  {author} {\bibinfo {author} {\bibfnamefont {C.}~\bibnamefont
  {Presilla}}, \bibinfo {author} {\bibfnamefont {R.}~\bibnamefont {Onofrio}},\
  and\ \bibinfo {author} {\bibfnamefont {U.}~\bibnamefont {Tambini}},\
  }\bibfield  {title} {\bibinfo {title} {Measurement quantum mechanics and
  experiments on quantum zeno effect},\ }\href@noop {} {\bibfield  {journal}
  {\bibinfo  {journal} {Annals of Physics}\ }\textbf {\bibinfo {volume}
  {248}},\ \bibinfo {pages} {95} (\bibinfo {year} {1996})}\BibitemShut
  {NoStop}%
\bibitem [{\citenamefont {Clausen}\ \emph {et~al.}(2010)\citenamefont
  {Clausen}, \citenamefont {Bensky},\ and\ \citenamefont
  {Kurizki}}]{addis36-clausen2010bath}%
  \BibitemOpen
  \bibfield  {author} {\bibinfo {author} {\bibfnamefont {J.}~\bibnamefont
  {Clausen}}, \bibinfo {author} {\bibfnamefont {G.}~\bibnamefont {Bensky}},\
  and\ \bibinfo {author} {\bibfnamefont {G.}~\bibnamefont {Kurizki}},\
  }\bibfield  {title} {\bibinfo {title} {Bath-optimized minimal-energy
  protection of quantum operations from decoherence},\ }\href@noop {}
  {\bibfield  {journal} {\bibinfo  {journal} {Physical Review Letters}\
  }\textbf {\bibinfo {volume} {104}},\ \bibinfo {pages} {040401} (\bibinfo
  {year} {2010})}\BibitemShut {NoStop}%
\bibitem [{\citenamefont {Hahn}(1950)}]{viola22-hahn1950spin}%
  \BibitemOpen
  \bibfield  {author} {\bibinfo {author} {\bibfnamefont {E.~L.}\ \bibnamefont
  {Hahn}},\ }\bibfield  {title} {\bibinfo {title} {Spin echoes},\ }\href@noop
  {} {\bibfield  {journal} {\bibinfo  {journal} {Physical review}\ }\textbf
  {\bibinfo {volume} {80}},\ \bibinfo {pages} {580} (\bibinfo {year}
  {1950})}\BibitemShut {NoStop}%
\bibitem [{\citenamefont {Viola}\ and\ \citenamefont
  {Lloyd}(1998{\natexlab{b}})}]{viola1998dynamical}%
  \BibitemOpen
  \bibfield  {author} {\bibinfo {author} {\bibfnamefont {L.}~\bibnamefont
  {Viola}}\ and\ \bibinfo {author} {\bibfnamefont {S.}~\bibnamefont {Lloyd}},\
  }\bibfield  {title} {\bibinfo {title} {Dynamical suppression of decoherence
  in two-state quantum systems},\ }\href@noop {} {\bibfield  {journal}
  {\bibinfo  {journal} {Physical Review A}\ }\textbf {\bibinfo {volume} {58}},\
  \bibinfo {pages} {2733} (\bibinfo {year} {1998}{\natexlab{b}})}\BibitemShut
  {NoStop}%
\bibitem [{\citenamefont {Uhrig}(2007)}]{uhrig2007keeping-addis60}%
  \BibitemOpen
  \bibfield  {author} {\bibinfo {author} {\bibfnamefont {G.~S.}\ \bibnamefont
  {Uhrig}},\ }\bibfield  {title} {\bibinfo {title} {Keeping a quantum bit alive
  by optimized $\pi$-pulse sequences},\ }\href@noop {} {\bibfield  {journal}
  {\bibinfo  {journal} {Physical Review Letters}\ }\textbf {\bibinfo {volume}
  {98}},\ \bibinfo {pages} {100504} (\bibinfo {year} {2007})}\BibitemShut
  {NoStop}%
\bibitem [{\citenamefont {Hodgson}\ \emph {et~al.}(2010)\citenamefont
  {Hodgson}, \citenamefont {Viola},\ and\ \citenamefont
  {D’Amico}}]{addis11-hodgson2010towards}%
  \BibitemOpen
  \bibfield  {author} {\bibinfo {author} {\bibfnamefont {T.~E.}\ \bibnamefont
  {Hodgson}}, \bibinfo {author} {\bibfnamefont {L.}~\bibnamefont {Viola}},\
  and\ \bibinfo {author} {\bibfnamefont {I.}~\bibnamefont {D’Amico}},\
  }\bibfield  {title} {\bibinfo {title} {Towards optimized suppression of
  dephasing in systems subject to pulse timing constraints},\ }\href@noop {}
  {\bibfield  {journal} {\bibinfo  {journal} {Physical Review A}\ }\textbf
  {\bibinfo {volume} {81}},\ \bibinfo {pages} {062321} (\bibinfo {year}
  {2010})}\BibitemShut {NoStop}%
\bibitem [{\citenamefont {Addis}\ \emph {et~al.}(2015)\citenamefont {Addis},
  \citenamefont {Ciccarello}, \citenamefont {Cascio}, \citenamefont {Palma},\
  and\ \citenamefont {Maniscalco}}]{addis2015dynamical-47song}%
  \BibitemOpen
  \bibfield  {author} {\bibinfo {author} {\bibfnamefont {C.}~\bibnamefont
  {Addis}}, \bibinfo {author} {\bibfnamefont {F.}~\bibnamefont {Ciccarello}},
  \bibinfo {author} {\bibfnamefont {M.}~\bibnamefont {Cascio}}, \bibinfo
  {author} {\bibfnamefont {G.~M.}\ \bibnamefont {Palma}},\ and\ \bibinfo
  {author} {\bibfnamefont {S.}~\bibnamefont {Maniscalco}},\ }\bibfield  {title}
  {\bibinfo {title} {Dynamical decoupling efficiency versus quantum
  non-markovianity},\ }\href@noop {} {\bibfield  {journal} {\bibinfo  {journal}
  {New Journal of Physics}\ }\textbf {\bibinfo {volume} {17}},\ \bibinfo
  {pages} {123004} (\bibinfo {year} {2015})}\BibitemShut {NoStop}%
\bibitem [{\citenamefont {Pokharel}\ \emph {et~al.}(2018)\citenamefont
  {Pokharel}, \citenamefont {Anand}, \citenamefont {Fortman},\ and\
  \citenamefont {Lidar}}]{pokharel2018demonstration}%
  \BibitemOpen
  \bibfield  {author} {\bibinfo {author} {\bibfnamefont {B.}~\bibnamefont
  {Pokharel}}, \bibinfo {author} {\bibfnamefont {N.}~\bibnamefont {Anand}},
  \bibinfo {author} {\bibfnamefont {B.}~\bibnamefont {Fortman}},\ and\ \bibinfo
  {author} {\bibfnamefont {D.~A.}\ \bibnamefont {Lidar}},\ }\bibfield  {title}
  {\bibinfo {title} {Demonstration of fidelity improvement using dynamical
  decoupling with superconducting qubits},\ }\href@noop {} {\bibfield
  {journal} {\bibinfo  {journal} {Physical Review Letters}\ }\textbf {\bibinfo
  {volume} {121}},\ \bibinfo {pages} {220502} (\bibinfo {year}
  {2018})}\BibitemShut {NoStop}%
\bibitem [{\citenamefont {Song}\ \emph {et~al.}(2017)\citenamefont {Song},
  \citenamefont {Tan},\ and\ \citenamefont {Kuang}}]{song2017control}%
  \BibitemOpen
  \bibfield  {author} {\bibinfo {author} {\bibfnamefont {Y.-J.}\ \bibnamefont
  {Song}}, \bibinfo {author} {\bibfnamefont {Q.-S.}\ \bibnamefont {Tan}},\ and\
  \bibinfo {author} {\bibfnamefont {L.-M.}\ \bibnamefont {Kuang}},\ }\bibfield
  {title} {\bibinfo {title} {Control quantum evolution speed of a single
  dephasing qubit for arbitrary initial states via periodic dynamical
  decoupling pulses},\ }\href@noop {} {\bibfield  {journal} {\bibinfo
  {journal} {Scientific reports}\ }\textbf {\bibinfo {volume} {7}},\ \bibinfo
  {pages} {1} (\bibinfo {year} {2017})}\BibitemShut {NoStop}%
\bibitem [{\citenamefont {Haseli}\ and\ \citenamefont
  {Salimi}(2020)}]{Haseli_2020}%
  \BibitemOpen
  \bibfield  {author} {\bibinfo {author} {\bibfnamefont {S.}~\bibnamefont
  {Haseli}}\ and\ \bibinfo {author} {\bibfnamefont {S.}~\bibnamefont
  {Salimi}},\ }\bibfield  {title} {\bibinfo {title} {Controlling the quantum
  speed limit time for unital maps via filtering operations},\ }\href
  {https://doi.org/10.1088/1612-202X/abac15} {\bibfield  {journal} {\bibinfo
  {journal} {Laser Physics Letters}\ }\textbf {\bibinfo {volume} {17}},\
  \bibinfo {pages} {105201} (\bibinfo {year} {2020})}\BibitemShut {NoStop}%
\bibitem [{\citenamefont {Leggett}\ \emph {et~al.}(1987)\citenamefont
  {Leggett}, \citenamefont {Chakravarty}, \citenamefont {Dorsey}, \citenamefont
  {Fisher}, \citenamefont {Garg},\ and\ \citenamefont
  {Zwerger}}]{addis28-leggett1987dynamics}%
  \BibitemOpen
  \bibfield  {author} {\bibinfo {author} {\bibfnamefont {A.~J.}\ \bibnamefont
  {Leggett}}, \bibinfo {author} {\bibfnamefont {S.}~\bibnamefont
  {Chakravarty}}, \bibinfo {author} {\bibfnamefont {A.~T.}\ \bibnamefont
  {Dorsey}}, \bibinfo {author} {\bibfnamefont {M.~P.}\ \bibnamefont {Fisher}},
  \bibinfo {author} {\bibfnamefont {A.}~\bibnamefont {Garg}},\ and\ \bibinfo
  {author} {\bibfnamefont {W.}~\bibnamefont {Zwerger}},\ }\bibfield  {title}
  {\bibinfo {title} {Dynamics of the dissipative two-state system},\
  }\href@noop {} {\bibfield  {journal} {\bibinfo  {journal} {Reviews of Modern
  Physics}\ }\textbf {\bibinfo {volume} {59}},\ \bibinfo {pages} {1} (\bibinfo
  {year} {1987})}\BibitemShut {NoStop}%
\bibitem [{\citenamefont {Chin}\ \emph
  {et~al.}(2012{\natexlab{a}})\citenamefont {Chin}, \citenamefont {Huelga},\
  and\ \citenamefont {Plenio}}]{song23-chin2012quantum}%
  \BibitemOpen
  \bibfield  {author} {\bibinfo {author} {\bibfnamefont {A.~W.}\ \bibnamefont
  {Chin}}, \bibinfo {author} {\bibfnamefont {S.~F.}\ \bibnamefont {Huelga}},\
  and\ \bibinfo {author} {\bibfnamefont {M.~B.}\ \bibnamefont {Plenio}},\
  }\bibfield  {title} {\bibinfo {title} {Quantum metrology in non-markovian
  environments},\ }\href@noop {} {\bibfield  {journal} {\bibinfo  {journal}
  {Physical Review Letters}\ }\textbf {\bibinfo {volume} {109}},\ \bibinfo
  {pages} {233601} (\bibinfo {year} {2012}{\natexlab{a}})}\BibitemShut
  {NoStop}%
\bibitem [{\citenamefont {Chin}\ \emph
  {et~al.}(2012{\natexlab{b}})\citenamefont {Chin}, \citenamefont {Huelga},\
  and\ \citenamefont {Plenio}}]{addis59-chin2012quantum}%
  \BibitemOpen
  \bibfield  {author} {\bibinfo {author} {\bibfnamefont {A.~W.}\ \bibnamefont
  {Chin}}, \bibinfo {author} {\bibfnamefont {S.~F.}\ \bibnamefont {Huelga}},\
  and\ \bibinfo {author} {\bibfnamefont {M.~B.}\ \bibnamefont {Plenio}},\
  }\bibfield  {title} {\bibinfo {title} {Quantum metrology in non-markovian
  environments},\ }\href@noop {} {\bibfield  {journal} {\bibinfo  {journal}
  {Physical Review Letters}\ }\textbf {\bibinfo {volume} {109}},\ \bibinfo
  {pages} {233601} (\bibinfo {year} {2012}{\natexlab{b}})}\BibitemShut
  {NoStop}%
\bibitem [{\citenamefont {Addis}\ \emph {et~al.}(2014)\citenamefont {Addis},
  \citenamefont {Brebner}, \citenamefont {Haikka},\ and\ \citenamefont
  {Maniscalco}}]{addis77-addis2014coherence}%
  \BibitemOpen
  \bibfield  {author} {\bibinfo {author} {\bibfnamefont {C.}~\bibnamefont
  {Addis}}, \bibinfo {author} {\bibfnamefont {G.}~\bibnamefont {Brebner}},
  \bibinfo {author} {\bibfnamefont {P.}~\bibnamefont {Haikka}},\ and\ \bibinfo
  {author} {\bibfnamefont {S.}~\bibnamefont {Maniscalco}},\ }\bibfield  {title}
  {\bibinfo {title} {Coherence trapping and information backflow in dephasing
  qubits},\ }\href@noop {} {\bibfield  {journal} {\bibinfo  {journal} {Physical
  Review A}\ }\textbf {\bibinfo {volume} {89}},\ \bibinfo {pages} {024101}
  (\bibinfo {year} {2014})}\BibitemShut {NoStop}%
\bibitem [{\citenamefont {Uhrig}(2008)}]{addis61-uhrig2008exact}%
  \BibitemOpen
  \bibfield  {author} {\bibinfo {author} {\bibfnamefont {G.~S.}\ \bibnamefont
  {Uhrig}},\ }\bibfield  {title} {\bibinfo {title} {Exact results on dynamical
  decoupling by $\pi$ pulses in quantum information processes},\ }\href@noop {}
  {\bibfield  {journal} {\bibinfo  {journal} {New Journal of Physics}\ }\textbf
  {\bibinfo {volume} {10}},\ \bibinfo {pages} {083024} (\bibinfo {year}
  {2008})}\BibitemShut {NoStop}%
\bibitem [{\citenamefont {Bellomo}\ \emph {et~al.}(2007)\citenamefont
  {Bellomo}, \citenamefont {Lo~Franco},\ and\ \citenamefont
  {Compagno}}]{Bellomo-PhysRevLett.99.160502}%
  \BibitemOpen
  \bibfield  {author} {\bibinfo {author} {\bibfnamefont {B.}~\bibnamefont
  {Bellomo}}, \bibinfo {author} {\bibfnamefont {R.}~\bibnamefont {Lo~Franco}},\
  and\ \bibinfo {author} {\bibfnamefont {G.}~\bibnamefont {Compagno}},\
  }\bibfield  {title} {\bibinfo {title} {Non-markovian effects on the dynamics
  of entanglement},\ }\href {https://doi.org/10.1103/PhysRevLett.99.160502}
  {\bibfield  {journal} {\bibinfo  {journal} {Physical Review Letters}\
  }\textbf {\bibinfo {volume} {99}},\ \bibinfo {pages} {160502} (\bibinfo
  {year} {2007})}\BibitemShut {NoStop}%
\bibitem [{\citenamefont {Xu}\ \emph {et~al.}(2014)\citenamefont {Xu},
  \citenamefont {Luo}, \citenamefont {Yang}, \citenamefont {Liu},\ and\
  \citenamefont {Zhu}}]{song13-xu2014quantum}%
  \BibitemOpen
  \bibfield  {author} {\bibinfo {author} {\bibfnamefont {Z.-Y.}\ \bibnamefont
  {Xu}}, \bibinfo {author} {\bibfnamefont {S.}~\bibnamefont {Luo}}, \bibinfo
  {author} {\bibfnamefont {W.}~\bibnamefont {Yang}}, \bibinfo {author}
  {\bibfnamefont {C.}~\bibnamefont {Liu}},\ and\ \bibinfo {author}
  {\bibfnamefont {S.}~\bibnamefont {Zhu}},\ }\bibfield  {title} {\bibinfo
  {title} {Quantum speedup in a memory environment},\ }\href@noop {} {\bibfield
   {journal} {\bibinfo  {journal} {Physical Review A}\ }\textbf {\bibinfo
  {volume} {89}},\ \bibinfo {pages} {012307} (\bibinfo {year}
  {2014})}\BibitemShut {NoStop}%
\bibitem [{\citenamefont {Liu}\ \emph {et~al.}(2015)\citenamefont {Liu},
  \citenamefont {Xu},\ and\ \citenamefont {Zhu}}]{song14-liu2015quantum}%
  \BibitemOpen
  \bibfield  {author} {\bibinfo {author} {\bibfnamefont {C.}~\bibnamefont
  {Liu}}, \bibinfo {author} {\bibfnamefont {Z.-Y.}\ \bibnamefont {Xu}},\ and\
  \bibinfo {author} {\bibfnamefont {S.}~\bibnamefont {Zhu}},\ }\bibfield
  {title} {\bibinfo {title} {Quantum-speed-limit time for multiqubit open
  systems},\ }\href@noop {} {\bibfield  {journal} {\bibinfo  {journal}
  {Physical Review A}\ }\textbf {\bibinfo {volume} {91}},\ \bibinfo {pages}
  {022102} (\bibinfo {year} {2015})}\BibitemShut {NoStop}%
\bibitem [{\citenamefont {Kheirandish}\ \emph {et~al.}(2008)\citenamefont
  {Kheirandish}, \citenamefont {Akhtarshenas},\ and\ \citenamefont
  {Mohammadi}}]{PhysRevA.77.042309}%
  \BibitemOpen
  \bibfield  {author} {\bibinfo {author} {\bibfnamefont {F.}~\bibnamefont
  {Kheirandish}}, \bibinfo {author} {\bibfnamefont {S.~J.}\ \bibnamefont
  {Akhtarshenas}},\ and\ \bibinfo {author} {\bibfnamefont {H.}~\bibnamefont
  {Mohammadi}},\ }\bibfield  {title} {\bibinfo {title} {Effect of spin-orbit
  interaction on entanglement of two-qubit heisenberg $xyz$ systems in an
  inhomogeneous magnetic field},\ }\href
  {https://doi.org/10.1103/PhysRevA.77.042309} {\bibfield  {journal} {\bibinfo
  {journal} {Physical Review A}\ }\textbf {\bibinfo {volume} {77}},\ \bibinfo
  {pages} {042309} (\bibinfo {year} {2008})}\BibitemShut {NoStop}%
\bibitem [{\citenamefont {Mohammadi}\ \emph {et~al.}(2011)\citenamefont
  {Mohammadi}, \citenamefont {Akhtarshenas},\ and\ \citenamefont
  {Kheirandish}}]{EPJD2011}%
  \BibitemOpen
  \bibfield  {author} {\bibinfo {author} {\bibfnamefont {H.}~\bibnamefont
  {Mohammadi}}, \bibinfo {author} {\bibfnamefont {S.}~\bibnamefont
  {Akhtarshenas}},\ and\ \bibinfo {author} {\bibfnamefont {F.}~\bibnamefont
  {Kheirandish}},\ }\bibfield  {title} {\bibinfo {title} {Influence of
  dephasing on the entanglement teleportation via a two-qubit heisenberg xyz
  system},\ }\href {https://doi.org/10.1140/epjd/e2011-10601-y} {\bibfield
  {journal} {\bibinfo  {journal} {European Physical Journal D}\ }\textbf
  {\bibinfo {volume} {62}},\ \bibinfo {pages} {439} (\bibinfo {year}
  {2011})}\BibitemShut {NoStop}%
\bibitem [{\citenamefont {Mohammadi}(2017)}]{QIP2017}%
  \BibitemOpen
  \bibfield  {author} {\bibinfo {author} {\bibfnamefont {H.}~\bibnamefont
  {Mohammadi}},\ }\bibfield  {title} {\bibinfo {title} {Post-markovian dynamics
  of quantum correlations: entanglement versus discord},\ }\href
  {https://doi.org/10.1007/s11128-016-1451-4} {\bibfield  {journal} {\bibinfo
  {journal} {Quantum Information Processing}\ }\textbf {\bibinfo {volume}
  {16}},\ \bibinfo {pages} {39} (\bibinfo {year} {2017})}\BibitemShut {NoStop}%
\bibitem [{\citenamefont {Milburn}(1991)}]{PhysRevA.44.5401}%
  \BibitemOpen
  \bibfield  {author} {\bibinfo {author} {\bibfnamefont {G.~J.}\ \bibnamefont
  {Milburn}},\ }\bibfield  {title} {\bibinfo {title} {Intrinsic decoherence in
  quantum mechanics},\ }\href {https://doi.org/10.1103/PhysRevA.44.5401}
  {\bibfield  {journal} {\bibinfo  {journal} {Physical Review A}\ }\textbf
  {\bibinfo {volume} {44}},\ \bibinfo {pages} {5401} (\bibinfo {year}
  {1991})}\BibitemShut {NoStop}%
\end{thebibliography}%

\end{document}